\documentclass[aps,twocolumn,groupedaddress,showpacs,showkeys,amsfonts,nofootinbib]{revtex4}

\usepackage{epsfig}
\usepackage{graphicx}
\usepackage{amsmath}

\newcommand{\m}{{{\hat m}_D}}
\newcommand{\C}{{\cal C}_{2}}
\newcommand{\LQCD}{{\Lambda_{\text{QCD}}}}

\newcommand{\tr}{{\rm tr}}

\newcommand{\bfA}{{{\boldsymbol{A}}}}
\newcommand{\bfk}{{{\boldsymbol{k}}}}

\newcommand{\bfx}{{{\boldsymbol{x}}}}
\newcommand{\bfy}{{{\boldsymbol{y}}}}

\newcommand{\PP}{{\rm P}}
\newcommand{\NP}{{\rm NP}}

\begin{document}

\title{
Power corrections in the quark-antiquark potential at finite temperature
}

\author{E. Meg\'{\i}as}
\email{emegias@ugr.es}

\author{E. \surname{Ruiz Arriola}}
\email{earriola@ugr.es}

\author{L.L. Salcedo}

\affiliation{
Departamento de F{\'\i}sica At\'omica, Molecular y Nuclear,
Universidad de Granada,
E-18071 Granada, Spain
}

\date{\today} 

\begin{abstract}
A  recently  proposed   phenomenological  model,  which  includes  nonperturbative effects from dimension  two gluon condensates, is applied
to analyze the available lattice  data for the heavy quark free energy
in  the  deconfined  phase  of  quenched QCD.   For  large  $q\bar{q}$
separations,  we  recover  previous  results  for  the  Polyakov  loop,
exhibiting  unequivocal power corrections in the temperature. For  the $q\bar{q}$
potential at finite  temperature and finite separation, we  find that a
good overall description of the  lattice data can be achieved once the
condensate is properly accounted  for. In addition, the model predicts
a duality between the zero  temperature potential as a function of the
$q\bar{q}$ separation, on the one  hand, and the quark selfenergy as a
function  of the  temperature, on  the other,  which turns  out  to be
satisfied to a high degree by the lattice data.
\end{abstract}

\pacs{11.10.Wx, 11.10.Kk, 11.15.-q, 11.10.Jj}

\keywords{ Polyakov Loop; Nonperturbative
Effects; Dimension two Condensate; QCD; Lattice QCD; Deconfinement
Phase; Thermal Field Theory}

\maketitle

\section{Introduction}
\label{sec:intro}

The understanding of the physics of the quark gluon plasma
\cite{McLerran:1986zb} has triggered a lot of activity in the past
years.  (See \cite{Adams:2005dq,Weiner:2005gp} for accounts of the
experimental situation.)  As a part of this effort, there have been
some recent advances in the study of the inter-quark forces at finite
temperature \cite{Mocsy:2004bv,Alberico:2006vw}.  In these analyses
the change in free energy due to the presence of a static
quark-antiquark pair separated by a distance~$r$ in a thermal bath has
often been used.  The color screening at high temperature will
produce a dissolution of the quarkonium state, and this could be a
signal of quark gluon plasma formation \cite{Matsui:1986dk}.

The vacuum  expectation value of the  Polyakov loop is  related to the
propagator of a static quark and constitutes a natural order parameter
for  the  deconfinement  phase  transition  of quenched  QCD.  Due  to
ultraviolet divergences  in the  heavy quark selfenergy,  the Polyakov
loop    is   subject    to    multiplicative   renormalization    (see
e.g.  \cite{KorthalsAltes:1993ca,Pisarski:2002ji}).  The  perturbative
computation of the renormalized Polyakov  loop has been carried out to
next to leading order \cite{Gava:1981qd} and only recently a practical
systematic procedure has been  devised to obtain it non perturbatively
in  lattice  QCD,  based  on  the computation  of  singlet  and  octet
correlation  functions  (Wilson lines)~\cite{Kaczmarek:2002mc}.  
These correlation functions are related to the free energy of a heavy
quark-antiquark pair.  In the limit of large separation the two
Polyakov loops decouple, and the result is essentially the square of
one single Polyakov loop. In the opposite limit of separations much
smaller than the Debye length the result is temperature independent,
so that the $q\bar{q}$ free energy reproduces the known QCD static
$q\bar{q}$ potential at zero temperature,
$V_{q\bar{q}}(r)$,~\cite{Necco:2001xg}.  There is an additive
ambiguity in the free energy that is fixed at short distances by
comparison with this zero temperature quantity. This strongly suggests
that there exists a deep connection between the zero temperature
$q\bar{q}$ potential and the vacuum expectation value of the Polyakov
loop.

The perturbative computation of the $q\bar{q}$ potential has been
studied for a long time. At tree-level, $V_{q\bar{q}}(r)$ is a Coulomb
potential arising from one-gluon exchange. Up to now there are
estimates of the three loop perturbative contribution using several
methods~\cite{Chishtie:2001mf,Pineda:2001zq,Cvetic:2003wk}. The
perturbative series turned out to be very poorly convergent at $r \ge
0.1 \, \textrm{fm}$ and resummation prescriptions lead to large
uncertainty due to scheme dependence. Several techniques have been
proposed to improve the perturbative series: extraction of the ${\cal
O}(\LQCD)$ infrared renormalons~\cite{Hoang:1998nz,Beneke:1998rk},
Operator Product Expansion (OPE) for $r \ll
\LQCD^{-1}$~\cite{Brambilla:1999qa}, etc.  Non-perturbative methods,
such as lattice simulations, and phenomenological potential model
analyses of experimental data for the heavy quarkonium spectra,
indicate that the $q\bar{q}$ potential can be well approximated by the
sum of a Coulomb term plus a linear term that becomes dominant at
separations of the order $\sim
1/\LQCD$~\cite{Bali:2000gf,Sumino:2003yp}. This non-perturbative
linear behavior at large distances is consistent with the quark
confinement picture based on the QCD string. For a recent attempt
within the holographic QCD approach based on the Ads/CFT
correspondence see e.g. \cite{Andreev:2006nw} and references therein.

On the other hand, at temperatures of the order of the deconfinement
phase transition the characteristic Euclidean scale corresponds to
$\mu \sim 4\pi T_c\sim 3\, \textrm{GeV}$, and one expects the OPE
formalism to apply~\cite{Wilson:1969zs}. This approach consists of
expanding the Green's functions in inverse power series of $p^2$, each
term corresponding to a local operator, implying that condensates and
power corrections could play a role. This has been verified in a
recent analysis of the renormalized Polyakov lattice
data~\cite{Kaczmarek:2002mc}, where the existence of non perturbative
contributions driven by gluon condensates has been
exposed~\cite{Megias:2005ve}. The first BRST invariant condensate is
of dimension two $\langle A_{\mu,a}^2\rangle$, with $A_{\mu,a}$ the
gluon field~\cite{Kondo:2001nq}. It is tempting to develop a model
based on the computation of correlations of Polyakov loops to connect
these two regimes (zero temperature and large distances), so that the
non-perturbative contributions in the Polyakov loop have the same
origin as those in the zero temperature $q\bar{q}$ potential, namely,
the gluon condensates. This will be one of the issues of the
manuscript. More generally, we extend the one-gluon exchange model
introduced in \cite{Megias:2005ve} for the Polyakov loop, to the
description of the singlet heavy quark-antiquark free energy and
compare with available lattice data. A preliminary study in this
direction was presented in \cite{Megias:2005pe}. 

The paper is organized as follows. In section~\ref{sec:2} we review
the observed non-perturbative behavior of the renormalized Polyakov
loop just above the deconfinement phase transition, and the importance
of thermal power corrections to describe it, thereby justifying the
introduction of dimension two condensates in the theory.  In
section~\ref{sec:3} we study the influence of these condensates on the
free energy of the static $q{\bar q}$ pair. In section~\ref{sec:4} we
compare our model with available lattice data and establish a
remarkable duality between the zero temperature $q{\bar q}$ potential
and the Polyakov loop.  Finally, in section~\ref{conclusions} we
summarize our points and draw our main conclusions.

\section{The Polyakov loop}
\label{sec:2}

In this section we summarize results derived for the Polyakov loop and
the role played by the dimension two condensate. Fuller details are
given in \cite{Megias:2005ve}.  The Polyakov loop or thermal Wilson
line is the propagator of a static color source.  Physically, the
expectation value of the $n$-point Polyakov loop correlation functions
are related to the change in free energy arising from the presence of
static quark and antiquark sources in the heat bath. The Polyakov loop
is defined by
\begin{equation}
\Omega(\bfx) = {\cal T} \exp \left( ig \int_0^{1/T} d x_0
  A_0(\bfx,x_0)\right) \,,
\label{eq:1}
\end{equation}
where ${\cal T}$ denotes time ordering, and $x_0$ is the Euclidean
time coordinate.  $A_0=\sum_a A_{0,a}T_a$ and $T_a$, $a=1,\ldots,
N_c^2-1$, are the generators of SU($N_c$) in the fundamental
representation, with normalization $\tr(T_a T_b)=\delta_{ab}/2$, and
$N_c$ the number of colors. Let
\begin{equation}
L(T)  = \left\langle \frac{1}{N_c}\tr\, \Omega \right\rangle \,,
\end{equation}
be the normalized expectation value of the Polyakov loop ($\tr$ is the
trace in the fundamental representation of the color group). This
quantity was computed perturbatively to next-to-leading order (NLO) by
Gava and Jengo \cite{Gava:1981qd}. To leading order (LO),
\begin{equation}
L(T)= 1+\frac{1}{16 \pi}\frac{N_c^2-1}{N_c}g^2\m  
+{\cal O}(g^4) \,,
\label{eq:2}
\end{equation}
where we have defined the dimensionless quantity $\m=m_D/T$, and $m_D$
is the Debye mass. To one loop \cite{Nadkarni:1983kb}
\begin{equation}
\m=\frac{m_D}{T} = g(N_c/3+N_f/6)^{1/2}\,
\label{eq:4}
\end{equation}
$N_f$ being the  number of flavors. The temperature  dependence in the
perturbative $L$ comes from the running of the coupling constant.

In a  static gauge  (i.e., one where  $A_0$ is time  independent), the
Polyakov loop takes the simpler form
\begin{equation}
\Omega(\bfx) = e^{ig A_0(\bfx)/T} \,,
\label{eq:3}
\end{equation}
and the computation of $L(T)$ can be undertaken using dimensional
reduction ideas
\cite{Ginsparg:1980ef,Appelquist:1981vg,Nadkarni:1983kb,Braaten:1996jr,Shaposhnikov:1996th}.
Namely, after the non stationary Matsubara modes are integrated out,
the static modes $A_0(\bfx)$ and $\bfA(\bfx)$ are described by an
effective three dimensional Lagrangian with computable
parameters.\footnote{A subtle point is that to make contact with the
Polyakov loop, the integration of non static Matsubara modes should be
done in a static gauge, whereas most calculations of the dimensionally
reduced Lagrangian are carried out in a covariant gauge. Fortunately
the spurious gauge dependence so generated only affects higher
perturbative orders \cite{Megias:2005ve}.} Expanding the exponential
one finds\footnote{Odd orders vanish assuming that the conjugation
symmetry $A_\mu(x)\to -A^T_\mu(x)$ is not spontaneously broken. By
the same token $\langle\Omega\rangle$ and
$\langle\Omega^\dagger\rangle$ are real (and equal in the absence of
baryonic chemical potential, which we assume throughout).}
\begin{equation}
L(T)=
1
-\frac{g^2}{2 T^2}\frac{1}{N_c}\langle\tr(A_0^2)\rangle
+\frac{g^4}{24 T^4}\frac{1}{N_c}\langle\tr(A_0^4)\rangle
+\cdots \,.
\end{equation}
The quartic term is ${\cal O}(g^6)$ and hence, to order ${\cal
O}(g^5)$ one can use the Gaussian-like approximation
\begin{equation}
 L  = \exp\left[-\frac{g^2\langle A_{0,a}^2\rangle}{4N_cT^2}
 \right]
+{\cal O}(g^6) \,.
\label{eq:7}
\end{equation}
This formula becomes exact in the large $N_c$ limit \cite{Megias:2006ke}. From
here it is immediate to relate the Polyakov loop to the gluon propagator
of the dimensionally reduced theory
\begin{equation}
\langle A_{0,a}(\bfx) A_{0,b}(\bfy)\rangle = \delta_{ab} T
\int \frac{d^3 k}{(2\pi)^3} e^{i\bfk\cdot
(\bfx-\bfy)}D_{00}(\bfk) \,.
\label{eq:8} 
\end{equation}
The lowest order perturbative term,
\begin{equation}
D^\PP_{00}(\bfk)=
\frac{1}{\bfk^2+m_D^2} + {\cal O}(g^2)\,,
\end{equation}
(the upperscript P refers to {\em perturbative} contribution) yields
\begin{equation}
\langle A_{0,a}^2\rangle^\PP = -(N_c^2-1)\frac{\m}{4\pi}T^2 
+ {\cal O}(g^2)\,,
\label{eq:10}
\end{equation}
where dimensional regularization rules have been applied to deal with
the UV divergence at coincident points. When combined with the
Gaussian-like formula (\ref{eq:7}) this term reproduces to LO the
perturbative result in (\ref{eq:2}). The NLO result is also reproduced
in this way \cite{Megias:2005ve}.

\begin{figure}[tbp]
\begin{center}
\epsfig{figure=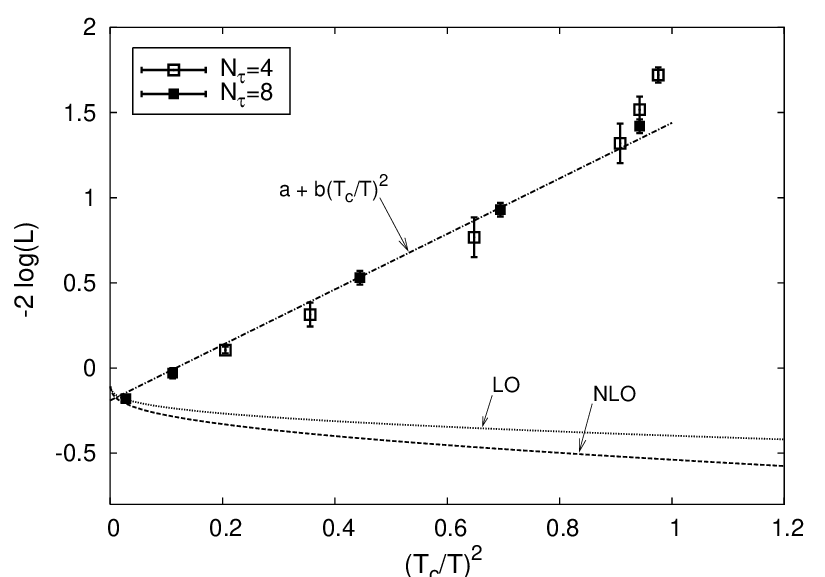,height=6.5cm,width=8.5cm}
\end{center}
\caption{The logarithm of the renormalized Polyakov loop in
 gluodynamics. Lattice data from \cite{Kaczmarek:2002mc} for
 $N_\tau=4$ (white squares) and $N_\tau=8$ (black squares). The
 straight line uses Eq.~(\ref{eq:11}) with the continuum extrapolated
 values of $a$ and $b$ in (\ref{eq:12}) (rather than the best fits to
 the $N_\tau=4$ or $N_\tau=8$ data). LO and NLO perturbative
 calculations \cite{Gava:1981qd} are shown for comparison.
}
\label{fig:1}
\end{figure}

Because in perturbation theory the dimensionless ratio $T/\LQCD$
appears only through logarithms from radiative corrections, the finite
temperature perturbative condensate $\langle
A_{0,a}^2\rangle^\PP$ is roughly proportional to $T^2$ and the
perturbative $L(T)$ is rather flat as a function of $T$. This is
definitively not what is seen in the lattice data, instead in the
unconfined phase one finds the pattern
\begin{eqnarray}
-2 \log{L} = a + b \left(\frac{T_c}{T}\right)^2 
\label{eq:11}
\end{eqnarray}
with $a$ and $b$ weakly dependent on $T$. This is shown in
Fig.~\ref{fig:1} for gluodynamics data
\cite{Kaczmarek:2002mc}\footnote{An alternative approach is taken in
\cite{Dumitru:2003hp} to measure the renormalized Polyakov loop in
lattice QCD.}. The same pattern applies for the unquenched data of
\cite{Kaczmarek:2005ui}. For later reference we quote the estimated
continuum limit of the fitted values of $a$ and $b$ for gluodynamics
\cite{Megias:2005ve}
\begin{equation}
a=-0.19(5),\quad b=1.63(16) \,.
\label{eq:12}
\end{equation}

A roughly constant $a$ can be accommodated with perturbation theory,
and this is so quantitatively in the high temperature region. The
second piece involves a power-like (as opposed to logarithmic)
temperature dependence of non perturbative origin, since $T_c$ is
proportional to $\LQCD$. On purely dimensional grounds it suggests a
temperature independent non perturbative dimension two condensate to
be added in $\langle A_{0,a}^2\rangle$. At zero temperature non
perturbative condensates appear naturally within the operator product
expansion approach \cite{Wilson:1969zs,Pascual:1984zb} through
modifications in the propagators of particles with respect to their
perturbative counterparts.  E.g.  the standard dimension four gluon
condensate \cite{Shifman:1978bx,Narison:1989aq} implies a term of the
type $\langle G_{\mu\nu}^2\rangle/\bfk^6$ to be added to the
perturbative gluon propagator $1/\bfk^2$.  Such non perturbative
addition is irrelevant in the very high momentum regime but introduces
sizeable power-like corrections as one approaches the strong
interacting and non perturbative low momentum regime. A dimension two
condensate has been proposed in the same zero temperature context in
\cite{Lavelle:1988eg,Chetyrkin:1998yr}.  This implies a gluon
propagator of the type $1/\bfk^2+m_G^2/\bfk^4$ which, on dimensional
grounds, yields a gluon exchange potential of the form Coulomb plus
string tension, $V\sim c/r+\sigma r$ (with $\sigma$ positive provided
$m_G^2$ is also positive, corresponding to a tachyonic gluon squared mass
$-m_G^2$).  As we will see immediately, such $m_G^2$ gives rise to a
dimension two condensate in the gluon field.

Guided by this insight the
same approach is taken in \cite{Megias:2005ve} to describe Polyakov
loop data at not so high temperatures above the deconfining
transition. That is, a non perturbative piece driven by a dimension
two operator is added in the finite temperature propagator:
\begin{equation}
D_{00}(\bfk)=D^\PP_{00}(\bfk)+D^\NP_{00}(\bfk)
\end{equation}
where
\begin{equation}
D^\NP_{00}(\bfk)= \frac{m_G^2}{(\bfk^2+m_D^2)^2} \,.
\end{equation}
This expression nicely combines the effects coming from high frequency
modes (the Debye mass from integration of the heavy thermal modes) and
from the low frequency modes incorporated in the condensates. Thus we
expect it to work in a window range above the phase transition.

Using (\ref{eq:8}) at $\bfx=\bfy$, the new term produces
\begin{equation}
\langle A_{0,a}^2\rangle= \langle A_{0,a}^2\rangle^\PP +
\langle A_{0,a}^2\rangle^\NP
\end{equation}
where the non perturbative condensate is related to $m_G^2$ through
\begin{equation}
\langle A_{0,a}^2\rangle^\NP = \frac{(N_c^2-1) m_G^2}{8\pi \m} \,.
\end{equation}
This contribution is UV finite. When inserted in (\ref{eq:7}) it
produces the term $b(T_c/T)^2$ in the pattern (\ref{eq:11}), assuming
that, $m_G^2$ or $\langle A_{0,a}^2\rangle^\NP$ are temperature
independent modulo radiative corrections. For future reference we
introduce the notation
\begin{equation}
\C := g^2\langle A_{0,a}^2\rangle^\NP
\end{equation}
to denote the non perturbative dimension two condensate. Remarkably,
the numerical value of the non perturbative finite temperature
condensate $\C$ turns out to be consistent \cite{Megias:2005ve} with
independent determinations of the zero temperature condensate
\cite{Boucaud:2001st,RuizArriola:2004en,Boucaud:2005rm}. This is in favor of the consistency of the
present approach.

\section{The singlet $Q{\bar Q}$ free energy}
\label{sec:3}

In this section we apply the previous model to the study of the free
energy of a heavy quark-antiquark pair. The latter is obtained through the
correlation of two Polyakov loops,
\begin{equation}
e^{-F_{q\bar{q}}(r,T)/T} = \left\langle \frac{1}{N_c}\tr\,\Omega(\bfx)
\frac{1}{N_c}\tr\,\Omega^\dagger(\bfy) \right\rangle , 
\label{eq:18}
\end{equation}
where $r=|\bfx-\bfy|$ is the separation between quark and
antiquark. The free energy contains the finite temperature $q{\bar q}$ potential as well
as the quarks selfenergies:
\begin{equation}
F_{q\bar{q}}(r,T) = V_{q\bar{q}}(r,T) + 2\, \Sigma_q(T),
\label{eq:19}
\end{equation}
where (twice) the selfenergy
\begin{equation}
2\Sigma_q(T)= \lim_{r\to\infty}F_{q\bar{q}}(r,T)
\end{equation}
is finite in the deconfined phase and in this case the potential
vanishes at large separations. In the confining phase the selfenergy
would be infinite and $F_{q\bar{q}}(r,T)$ is directly identified with
the $q\bar{q}$ potential $V_{q\bar{q}}(r,T)$. Of course, at large
separations the two Polyakov loops in (\ref{eq:18}) become
uncorrelated implying that the quark (or antiquark) selfenergy is
related to the expectation value of the Polyakov loop through
\begin{equation}
L(T)= e^{-\Sigma_q(T)/T}.
\end{equation}

Because the interaction between the color singlets $\tr\,\Omega$ and
$\tr\,\Omega^\dagger$ requires the exchange of at least two gluons
(each carrying a Debye screening factor $e^{-m_D r}$ at large
separations), this so called {\em color averaged} free energy is less suited
to lattice computation than the {\em singlet} free energy
\begin{equation}
e^{-F_1(r,T)/T} = \left\langle
\frac{1}{N_c}\tr\left(\Omega(\bfx)\Omega^\dagger(\bfy)\right) \right\rangle .
\end{equation}
This quantity defined in the Coulomb gauge acquires a gauge invariant
meaning \cite{Philipsen:2002az} and lattice calculations for it are
available in the literature \cite{Kaczmarek:2004gv}. The corresponding
$q{\bar q}$ potential is dominated by a single gluon exchange falloff
at large separations.

Following precisely the same approach described in the previous
section for the Polyakov loop, we obtain the relation analogous to
(\ref{eq:7})
\begin{equation}
\left\langle
\frac{1}{N_c}\tr\left(\Omega(\bfx)\Omega^\dagger(\bfy)\right) \right\rangle = 
e^{\frac{g^2}{2N_cT^2}\left( \langle
A_{0,a}(\bfx)A_{0,a}(\bfy)\rangle -\langle A_{0,a}^2\rangle \right)
}
\end{equation}
modulo ${\cal O}(g^6)$ corrections. Once again, based on the cumulant
expansion \cite{Megias:2006ke} these corrections vanish in the large
$N_c$ limit. From the relation $F_1=V_1+2\Sigma_q$ similar to
(\ref{eq:19})\footnote{Due to SU($N_c$) invariance,
$\langle\Omega\rangle$ is a multiple of the identity matrix and so
$\Sigma_q$ is the same quantity in $F_{q\bar{q}}$ and $F_1$.} we
identify the two contributions
\begin{eqnarray}
V_1(r,T) &=& -\frac{g^2}{2N_c T} \langle A_{0,a}(\bfx)A_{0,a}(\bfy)\rangle ,
\nonumber \\
\Sigma_q(T) &=& \frac{g^2}{4N_c T}\langle A_{0,a}^2\rangle,
\end{eqnarray}
and in particular the results of the previous section for the Polyakov
loop expectation value are recovered with (at LO)
\begin{equation}
\Sigma_q(T)= \frac{g^2 (N_c^2-1)}{16\pi N_c}\left(-\m T
+\frac{m_G^2}{2\m}\frac{1}{T}\right).
\label{eq:23}
\end{equation}
For the color singlet potential we derive the expression (at LO and
neglecting any scale dependence in the coupling constant)
\begin{equation}
V_1(r,T)= -\frac{g^2 (N_c^2-1)}{8\pi N_c}\left(\frac{1}{r}
+\frac{m_G^2}{2\m}\frac{1}{T}\right) e^{-\m T r}.
\label{eq:24}
\end{equation}

In these results the Polyakov loop is already renormalized by
dimensional regularization. In the corresponding lattice computation
the bare Polyakov loop has to be renormalized using a suitable
prescription. Specifically, in \cite{Kaczmarek:2002mc,Kaczmarek:2004gv}
the following criterion is used: the multiplicative renormalization of
the Polyakov loop translates into an additive renormalization in the
$q\bar{q}$ free energy,
\begin{equation}
F_1^{\text{renorm}}(r,T)=F_1^{\text{bare}}(r,T) +c(T),
\end{equation}
therefore, because at short distances the renormalized free energy
ought to be temperature independent, a suitable $r$-independent (but
$T$-dependent) constant $c(T)$ is added in such a way that at short
distances the (renormalized) finite temperature free energy $F_1(r,T)$
has a maximum overlap with the zero temperature one $F_1(r,0)$. The
contact between the universal zero temperature curve and the
temperature dependent one occurs for all separations below a distance of
the order of the Debye length \cite{Kaczmarek:2002mc}. This device
allows a quite accurate lattice determination of the selfenergy, and
hence of the renormalized Polyakov loop. In practical terms the above
criterion means that, if the free energy is expanded in powers of the
separation,
\begin{equation}
F_1(r,T)= v_0/r+v_1+v_2 r + {\cal O}(r^2)
\end{equation}
(neglecting the logarithmic dependence from the running of the
coupling constant), the constant term $v_1$ should be absent (i.e.
the Cornell prescription). This is because the dominant Coulombian
term $v_0/r$ is temperature independent, and $v_1$ vanishes in
$F_1(r,0)$ which is of the Coulomb plus string tension form. (A
non-vanishing $v_1$ would require an inexistent ``dimension one''
condensate.)

As it turns out, our formulas (\ref{eq:23}) and (\ref{eq:24})
automatically satisfy this Cornell-Bielefeld prescription for the free
energy. I.e., in $F_1=V_1+2\Sigma_q$ the term $v_1$ is already zero
without any further renormalization. This is most remarkable since the
selfenergy used a dimensionally regularized definition of the UV
divergent integral $\int d^3 k\, D^\PP_{00}(\bfk)$ leading to
(\ref{eq:10}), whereas the potential is UV finite for any non
vanishing value of $r$. Actually, this was to be expected because, as
is well-known, dimensional regularization does not introduce any
power-like dependence in addition to those already existing in the
theory, so a dimensionful $v_1$ could not be introduced in the
calculation. By the same token we expect that the Cornell-Bielefeld
prescription will be automatically satisfied also at higher orders,
i.e., beyond the Gaussian-like approximation, provided dimensional
regularization is used. Note also that the fact that the
Cornell-Bielefeld prescription used in lattice is satisfied by our
approach justifies the comparison of our calculation of the
renormalized Polyakov loop, based on single loops, with that in
lattice, based on the correlation of pairs of loops.

\section{Comparison with lattice data}
\label{sec:4}

\subsection{Duality between the zero temperature $Q\bar{Q}$-potential and the Polyakov loop}
\label{sec:4A}

In this subsection we will analyze within our model the $q\bar{q}$
singlet free energy with respect to the two asymptotic behaviors of
zero temperature (at finite separation) and large separation (at
finite temperature). Neglecting momentarily any scale dependence in
both the coupling constant and the non perturbative condensate, we
obtain (at LO)
\begin{equation}
F_1(\infty,T)= -\frac{ (N_c^2-1)}{8\pi N_c}g^2\m T
+ \frac{ \C}{2 N_c}\frac{1}{T}
\label{eq:27}
\end{equation}
for the singlet free energy at infinite separation, and
\begin{equation}
F_1(r,0)= -\frac{ (N_c^2-1)}{2N_c}\frac{g^2}{4\pi}\frac{1}{r}
+\frac{\C}{2N_c}  \m r
\label{eq:28}
\end{equation}
for the free energy at zero temperature (i.e.,
$V_{q\bar{q}}(r)$)\footnote{Although the model is devised for the
deconfined phase, nothing prevents us from taking the formal zero
temperature limit in our expression for the free energy.}. The first
thing to observe is that the latter takes the standard zero
temperature form of a perturbative Coulombian term plus a linearly
raising confining term $\sigma r$, with string tension\footnote{At
finite temperature, the ${\cal O}(r)$ piece of the free energy is
proportional to $m_G^2-m_D^2$ indicating a weakening of the string
tension. Also it confirms that the size of the contact between the
zero and finite temperature curves is of the order of $1/m_D$.}
\begin{equation}
\sigma= \frac{1}{2N_c}\C \m \,.
\label{eq:29}
\end{equation}

On the other hand, another obvious property in (\ref{eq:27}) and
(\ref{eq:28}) is the formal resemblance between the two asymptotic
free energies with the identification $r\sim 1/T$. More precisely
\begin{equation}
F_1(\infty,T)= F_1(r=1/m_D,0).
\label{eq:30}
\end{equation}
This relation indicates a remarkable and striking duality, within our
model, between the quark selfenergy and the zero temperature
$q\bar{q}$ potential. Such symmetry can be traced to the form of the
propagator $D_{00}(\bfk;T)$, namely, at least at LO its dependence on
momentum and temperature comes in the combination $\bfk^2+m_D^2$,
regardless of the value of the coupling constant and the non
perturbative condensate. Unfortunately the data are not available in
momentum space. This makes difficult to check the scaling of the two
point function directly. However, when the momentum $\bfk$ is traded
by the separation $r$, this symmetry can still be seen as a duality
between the two asymptotic quantities $F_1(\infty,T)$ and $F_1(r,0)$.

Of course the abovementioned duality between the roles played by
$\bfk$ and $m_D$ will be disturbed by any scale dependence in the
coupling constant and in the condensate corresponding to scaling
violations. Such scale dependence implies a possible $r$ and $T$
dependence in those quantities and will be studied further below. Here
we take $g$ and $\C$ as constant (scale independent) but admit for
them the possibility of different values in the two asymptotic regimes
$(r,T=0)$ and $(r=\infty,T)$. For the coupling constant this
assumption is sustained by the data: the bosonic string model fits
extremely well the zero temperature $q\bar{q}$ potential, with an
error at the level of the $1\%$ in the interval $0.15\,\textrm{fm} \le
r \le 0.8\,\textrm{fm}$ \cite{Necco:2001xg}. In this model the
coupling constant takes a fixed value
\begin{equation}
g(r,0)= g_r:=\frac{\pi}{2}\,.
\label{eq:31}
\end{equation}
Of course asymptotic freedom requires $g(r,0)$ to go to zero as $r$
decreases, however this is only visible for separations below
$0.1\,\textrm{fm}$ \cite{Kaczmarek:2004gv}.\footnote{Actually, the
coupling $g(\mu)=\pi/2$ matching the string bosonic model
corresponds to a NLO evolution scale $1/\mu \sim 0.1 \,\text{fm}$ for 
$\LQCD\sim 240\,\text{MeV}$.}  On the other hand,
in the large separation regime $(r=\infty,T)$, the Polyakov loop data
favor a constant $g$ (i.e. a constant $a$ in (\ref{eq:11})) rather
than the perturbative running one. The constancy of $a$ is unveiled
only after the power correction has been subtracted off. Once again,
asymptotic freedom requires $g(\infty,T)$ to decrease as $T$ increases
but this perturbative regime occurs at very high temperatures
only. This ``freezing'' of $g(\infty,T)$ to a constant value at
intermediate temperatures is more thoroughly analyzed from the
$q\bar{q}$ potential lattice data in the next subsection. So we take
\begin{equation}
g(\infty,T)= g_T
\end{equation}
with $g_T$ obtained from the relation
\begin{equation}
a=-\frac{N_c^2-1}{2N_c}\frac{g^2}{4\pi}\m\,.
\end{equation}
For $N_c=3$, $N_f=0$ and $a=-0.19$ this gives 
\begin{equation}
g_T=1.21(11) \,.
\label{eq:34}
\end{equation}
The fact that both $g_r$ and $g_T$ become frozen at intermediates
scales in the two regimes is in support of the duality discussed
above.

Let us now turn to the condensate, to see whether the relation
(\ref{eq:29}) between the string tension and the condensate is
verified. From the gluodynamics value of $b$ in (\ref{eq:12}) one
obtains a condensate
\begin{equation}
\C = (0.84(4)\,\textrm{GeV})^2 \,.
\label{eq:36}
\end{equation}
When this is combined in (\ref{eq:29}) with $g=g_r$ (needed in $\m$), it yields
\begin{equation}
\sigma= (0.43(2)\,\textrm{GeV})^2 \,.
\end{equation}
Remarkably, this is consistent with the accepted value of the string
tension $\sigma \approx (0.42\,\textrm{GeV})^2$ extracted, e.g., from
the slope of the Regge trajectories \cite{Anisovich:2000kx}. In view
of this, we will assume that the non perturbative condensate takes a
common value in the two asymptotic regimes $(r,T=0)$ and
$(r=\infty,T)$.

Taking into account that $g$, but not the condensate, takes two
different values in the two regimes, the duality relation
(\ref{eq:30}) gets modified since one has to use $g$ equal to $g_T$ in
(\ref{eq:27}) and equal to $g_r$ in (\ref{eq:28}).\footnote{Including
the $g$ in $\m$, but of course not that in $\C$.} This yields the
refined duality relation
\begin{equation}
F_1(\infty,T)= \gamma^{-3} F_1(r=\gamma/m_D,0)
\label{eq:37}
\end{equation}
where
\begin{equation}
\gamma = \left(\frac{g_r}{g_T}\right)^{1/2} \,, \quad
m_D = g_T (N_c/3 + N_f/6)^{1/2}T \,. 
\label{eq:38}
\end{equation}

\begin{figure}[tbp]
\begin{center}
\resizebox{0.5\textwidth}{!}{%
 \includegraphics{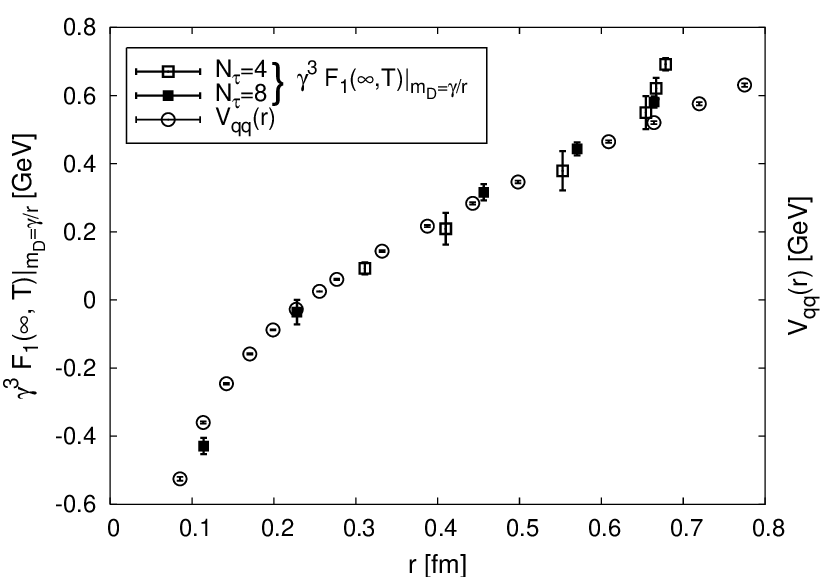}
}
\end{center}
\caption{ Verification of scaling relations in (\ref{eq:37}). Black
and white squares: $\gamma^3 F_1(\infty,T)$ from lattice data in
\cite{Kaczmarek:2002mc}. Circles: lattice data for the zero
temperature $q\bar{q}$ potential from \cite{Necco:2001xg}. Both
quantities are plotted as functions of the separation $r$ using the
relation $r\leftrightarrow \gamma/m_D$.}
\label{fig:2}
\end{figure}

\begin{figure}[tbp]
\begin{center}
\resizebox{0.5\textwidth}{!}{%
 \includegraphics{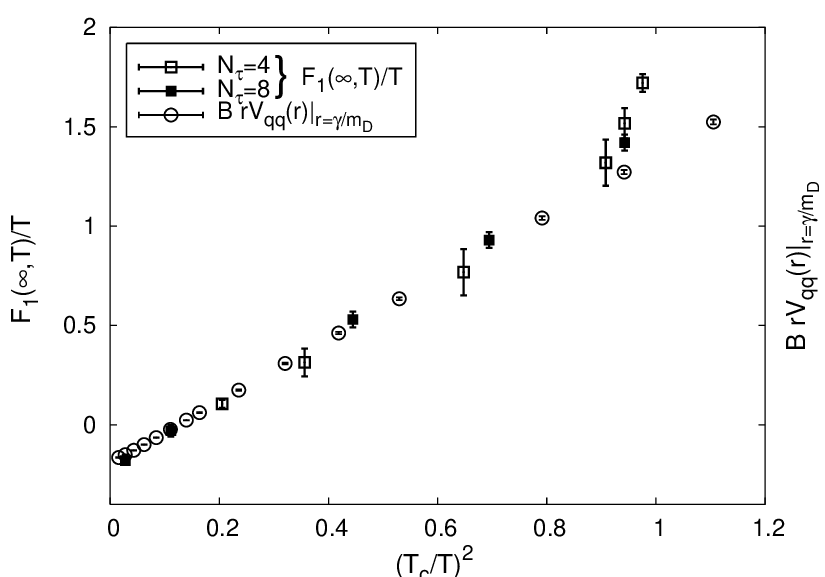}
}
\end{center}
\caption{Same lattice data as in Fig.~\ref{fig:2} for the
  dimensionless quantities $F_1(\infty,T)/T$ (squares) and
  $BrV_{q\bar{q}}(r)$ (circles), as functions of $(T_c/T)^2$.
  ($B=\m\gamma^{-4}$.) Both the duality and the straight line pattern are
  exhibited by the data.}
\label{fig:3}
\end{figure}

To see how this relation works, we plot in Fig.~\ref{fig:2} lattice
data points available for the two quantities: on the one hand twice
the quark selfenergy, $F_1(\infty,T)$ from \cite{Kaczmarek:2002mc},
and on the other the zero temperature $q\bar{q}$ potential $F_1(r,0)$
from \cite{Necco:2001xg}.  To compare both quantities we use the
relation $r\leftrightarrow\gamma/m_D$ and the rescaling factor
$\gamma^3$ indicated in (\ref{eq:37}).  $\gamma$ is computed using the
numerical values of $g_r$ and $g_T$ quoted in (\ref{eq:31}) and
(\ref{eq:34}), which correspond to a continuum extrapolation of those
parameters.\footnote{Note that the lattice data for $F_1(r,0)$, unlike
those of $F_1(\infty,T)$, have been extrapolated to the continuum.}
The remarkable agreement (modulo simple rescaling) between the two
seemingly unrelated quantities suggests that this duality might
actually be a feature of QCD, uncovered by our model. (Of course, any
duality should break down at temperatures below the transition, since
nothing is expected to happen to the zero temperature $q\bar{q}$
potential for separations above $1/T_c$.) The same data are shown in
Fig.~\ref{fig:3}, using the dimensionless quantities $F_1(\infty,T)/T$
and (scaled) $r F_1(r,0)$, as functions of $(T_c/T)^2$.  The straight
line patterns (\ref{eq:27}) and (\ref{eq:28}) are clearly exhibited,
indicating the dominant role played by a non perturbative dimension
two operator in the data.

\subsection{Singlet free energy at finite $r$ and $T$}
\label{sec:4B}

In order to further check our model against non perturbative
information we will use the high quality lattice data of
\cite{Kaczmarek:2004gv}, where the singlet $q\bar{q}$ free energy is
computed for quenched QCD ($N_c=3$) at various temperatures in the
confining phase for temporal sizes $N_\tau=4,8$.

In order to make a comparison we use the following parameterization of
the potential
\begin{equation}
V_1(r,T)= -\left(\frac{g^2}{3\pi}\frac{1}{r} +
\frac{\C}{2N_c T} \right) e^{-\m T r}
\end{equation}
where $\m$ is related to $g$ as in (\ref{eq:4}) with $N_c=3$,
$N_f=0$. Furthermore, we take an $r$-dependent $g$ consistent with the
zero temperature $q\bar{q}$-potential in \cite{Necco:2001xg}, namely,
for each $r$, $g(r)$ is taken so that
\begin{equation}
V_{q\bar{q}}(r)=-\frac{g^2(r)}{3\pi r} + \sigma r.
\end{equation}
\begin{figure}[tbp]
\begin{center}
\resizebox{0.5\textwidth}{!}{%
\includegraphics{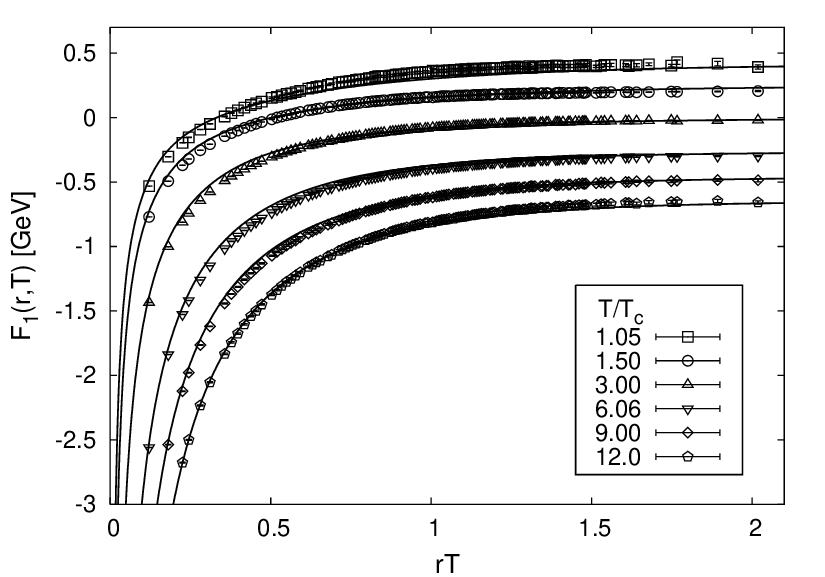}
}
\end{center}
\caption{Points: lattice data, for $N_\tau=8$, for the singlet free
energy as a function of $rT$, at $T/T_c=1.05,1.5,3,6.06,9,$ and $12$
(from top to bottom, respectively) from Ref.~\cite{Kaczmarek:2004gv}.
Curves: $F_1(r,T)=V_1(r,T)+F_1(\infty,T)$ of our model.  $g(r)$ (to be
used in $V_1(r,T)$) is extracted from the zero temperature potential
\cite{Necco:2001xg}.  In addition $C_2=(0.9\,\text{GeV})^2$, and
$g_T=1.26$.}
\label{fig:4}
\end{figure}
Following \cite{Necco:2001xg}, for $r$ below $0.1\,\text{fm}$ the
perturbative value of $V_{q\bar{q}}(r)$ to three-loops is used, and
above $0.1\,\text{fm}$, $g(r)=\pi/2$ from the bosonic string model. As
noted before this hybrid model reproduces remarkably well the lattice
data.  The results are presented in Fig.~\ref{fig:4} where the model
is compared with $N_\tau=8$ lattice data (the results for $N_\tau=4$
data are qualitatively similar). For clarity we represent
$F_1$ instead of $V_1$, and use $rT$ as the independent variable.  For
the temperature dependent shift $F_1(\infty,T)$ the expression
(\ref{eq:27}) is used with $g_T=1.26$, which corresponds to a value
$a=-0.21$; this is the LO perturbative value around $T=6\,T_c$. For
the condensate we take $\C=(0.9\,\text{GeV})^2$ which gives a fair
account of $F_1(\infty,T)$ at lower temperatures. (Note that the
points in the figure are the actual lattice data, without any
continuum extrapolation.)  For consistency the same value of $\C$ is
used in $V_1(r,T)$ but the precise value is not very important in the
potential (also it is not very important for $F_1(\infty,T)$ for high
temperatures). As we can see from Fig.~\ref{fig:4}, the overall
pattern of the lattice data is reasonably well reproduced by our
model. Given the small error bars in the data, a more detailed
agreement (small $\chi^2/\text{pdof}$) was not to be expected without
further sophistication of the model. As was noted in Section
\ref{sec:2}, $F_1(\infty,T)$ (or equivalently, the Polyakov loop)
cannot be described at all without introducing a dimension two
condensate.  The same holds here for the potential: the overall shape
of lattice data for $V_1(r,T)$ cannot be reproduced even approximately
if $\C$ is set to zero, moreover, the same value of the condensate
extracted form $F_1(\infty,T)$ works for the potential too.  This
feature is expected to survive in more elaborated models.

\begin{figure}[tbp]
\begin{center}
\resizebox{0.5\textwidth}{!}{%
\includegraphics{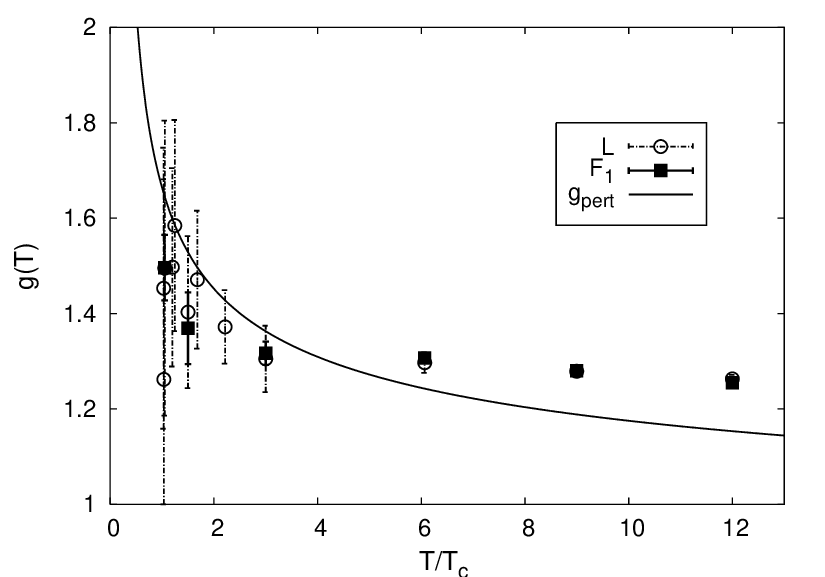}
}
\end{center}
\caption{Points: values of $g_T$ extracted from a fit to the Polyakov
loop data \cite{Kaczmarek:2002mc,Zantow:2003uh} (circles) and from $F_1(r,T)$
lattice data \cite{Kaczmarek:2004gv} (squares). LO formulas have been
used and $C_2=(0.90(5)\,\text{GeV})^2$. The curve displays the
perturbative running coupling constant at three-loops.
}
\label{fig:5}
\end{figure}
We have considered also a fit to the data allowing a $T$ dependent
$g_T$ (but keeping $\C$ constant).  This is depicted in
Fig.~\ref{fig:5}.  The fit using all data for the free energy is of
course more stable than that using only the Polyakov loop, although
both sets yield consistent results. The error bars are very small at
high temperatures and come from the lattice data and from allowing a
variation range $\C=(0.90\pm 0.05\,\text{GeV})^2$ in the condensate.
We find an almost flat dependence of $g_T$ with the temperature, much
flatter than expected from a perturbative running. As noted, a nearly
constant $g_T$ is in favor of the duality relation displayed in
Fig.~\ref{fig:2} and ~\ref{fig:3}, since $g_r$ is also frozen in the
bosonic string model. $g_T$ is somewhat larger than the perturbative
value. It is noteworthy that a larger renormalization of the coupling
constant, of about $50\%$, has also been observed in
\cite{Kaczmarek:2004gv,Kaczmarek:2005jy} if the observed Debye mass,
extracted from the exponential falloff of the potential, is compared
to its perturbative value. In our model the exponential fall at large
separations in controlled by $g_r$ rather than $g_T$. Since
$g_r=\pi/2$ is rather large, as compared to the perturbative value,
this is qualitative agreement with the findings in
\cite{Kaczmarek:2004gv,Kaczmarek:2005jy} (see also
\cite{Agasian:2006ra}), however, $g_r$ is somewhat smaller and does
not run.

The present analysis of the singlet $q\bar{q}$ free energy is carried
out at LO, and further studies will have to take into account higher
orders in the loop expansion. We certainly do not expect these higher
order corrections to reproduce the effect of condensates. In addition,
corrections coming from the condensate anomalous dimension are also
expected to be relevant in more accurate descriptions; at the present
stage this effect on the non-perturbative side should play a
comparable role to loops corrections. Finally, higher dimensional
condensates could also be present and affect the free energy.
Unfortunately, as our previous analysis reveals~\cite{Megias:2005ve},
the extraction of a dimension four condensate from the lattice data of
the Polyakov loop is not reliable at present. It would contribute to
(\ref{eq:11}) with a term $(T_c/T)^4$.

\section{Conclusions}
\label{conclusions}

In the present paper we have investigated the role of dimension two
condensates in the interrelation between the Polyakov loop at finite
temperature above the phase transition and the static quark-antiquark
potential at zero temperature. This extends our previous model for the
renormalized Polyakov loop in the deconfined phase which proved very
successful and enabled a quantitative determination of the BRST
invariant dimension two gluon condensate~\cite{Megias:2005ve}. Here we
have studied the consequences of the model when applied to the heavy
quark-antiquark free energy in the unconfined phase of QCD.

In our approach the perturbative and non-perturbative contributions
are explicitly separated and an analytic and simple expression for the
free energy which interpolates between the short and long distances
regime is derived.  The separation turns out to accommodate quite
naturally a smooth running coupling when the model is confronted to
lattice data, which speaks in favor of the consistency of the
approach.

We observe a natural consequence of the presence of the dimension two
condensate if one restricts to LO in the perturbative contributions
and effects associated to anomalous dimensions are
disregarded. Although in general the free energy of a quark-antiquark
pair depends both on the distance $r$ and the temperature $T$, there
are some regions where this double dependence presents strong
correlations. Amazingly, the regimes of finite temperature and long
distances (from which the Polyakov loop is defined) and the zero
temperature and finite distances (from which the zero temperature
$q\bar{q}$ potential is obtained) have a striking dual functional
dependence which is verified on a quantitative level on the lattice if
the simple change of variable $r \leftrightarrow 1/m_D$ is assumed.
The very fact that the non-perturbative contributions are encoded in
the dimension two condensate suggests that our result holds up to mild
logarithmic radiative corrections and it would be interesting to
quantify the observed small discrepancies in a more systematic manner.
Despite all these caveats we believe that the quantitative success of
the duality relations provides a further evidence on existence and
relevance of dimension two gluon condensates.

\begin{acknowledgments}
We thank O. Kaczmarek and F. Karsch for kindly providing us with the
lattice data of Ref. \cite{Kaczmarek:2004gv}. This work is supported
in part by funds provided by the Spanish DGI and FEDER funds with
grant no. FIS2005-00810, Junta de Andaluc{\'\i}a grants no.  FQM225-05
and FQM481 and EU Integrated Infrastructure Initiative Hadron Physics
Project contract no. RII3-CT-2004-506078.
\end{acknowledgments}


\begin{thebibliography}{46}
\expandafter\ifx\csname natexlab\endcsname\relax\def\natexlab#1{#1}\fi
\expandafter\ifx\csname bibnamefont\endcsname\relax
  \def\bibnamefont#1{#1}\fi
\expandafter\ifx\csname bibfnamefont\endcsname\relax
  \def\bibfnamefont#1{#1}\fi
\expandafter\ifx\csname citenamefont\endcsname\relax
  \def\citenamefont#1{#1}\fi
\expandafter\ifx\csname url\endcsname\relax
  \def\url#1{\texttt{#1}}\fi
\expandafter\ifx\csname urlprefix\endcsname\relax\def\urlprefix{URL }\fi
\providecommand{\bibinfo}[2]{#2}
\providecommand{\eprint}[2][]{\url{#2}}

\bibitem[{\citenamefont{McLerran}(1986)}]{McLerran:1986zb}
\bibinfo{author}{\bibfnamefont{L.~D.} \bibnamefont{McLerran}},
  \bibinfo{journal}{Rev. Mod. Phys.} \textbf{\bibinfo{volume}{58}},
  \bibinfo{pages}{1021} (\bibinfo{year}{1986}).

\bibitem[{\citenamefont{Adams et~al.}(2005)}]{Adams:2005dq}
\bibinfo{author}{\bibfnamefont{J.}~\bibnamefont{Adams}} \bibnamefont{et~al.}
  (\bibinfo{collaboration}{STAR}), \bibinfo{journal}{Nucl. Phys.}
  \textbf{\bibinfo{volume}{A757}}, \bibinfo{pages}{102} (\bibinfo{year}{2005}),
  \eprint{nucl-ex/0501009}.

\bibitem[{\citenamefont{Weiner}(2006)}]{Weiner:2005gp}
\bibinfo{author}{\bibfnamefont{R.~M.} \bibnamefont{Weiner}},
  \bibinfo{journal}{Int. J. Mod. Phys.} \textbf{\bibinfo{volume}{E15}},
  \bibinfo{pages}{37} (\bibinfo{year}{2006}), \eprint{hep-ph/0507115}.

\bibitem[{\citenamefont{Mocsy and Petreczky}(2005)}]{Mocsy:2004bv}
\bibinfo{author}{\bibfnamefont{A.}~\bibnamefont{Mocsy}} \bibnamefont{and}
  \bibinfo{author}{\bibfnamefont{P.}~\bibnamefont{Petreczky}},
  \bibinfo{journal}{Eur. Phys. J.} \textbf{\bibinfo{volume}{C43}},
  \bibinfo{pages}{77} (\bibinfo{year}{2005}), \eprint{hep-ph/0411262}.

\bibitem[{\citenamefont{Alberico et~al.}(2006)\citenamefont{Alberico, Beraudo,
  De~Pace, and Molinari}}]{Alberico:2006vw}
\bibinfo{author}{\bibfnamefont{W.~M.} \bibnamefont{Alberico}},
  \bibinfo{author}{\bibfnamefont{A.}~\bibnamefont{Beraudo}},
  \bibinfo{author}{\bibfnamefont{A.}~\bibnamefont{De~Pace}}, \bibnamefont{and}
  \bibinfo{author}{\bibfnamefont{A.}~\bibnamefont{Molinari}}
  (\bibinfo{year}{2006}), \eprint{hep-ph/0612062}.

\bibitem[{\citenamefont{Matsui and Satz}(1986)}]{Matsui:1986dk}
\bibinfo{author}{\bibfnamefont{T.}~\bibnamefont{Matsui}} \bibnamefont{and}
  \bibinfo{author}{\bibfnamefont{H.}~\bibnamefont{Satz}},
  \bibinfo{journal}{Phys. Lett.} \textbf{\bibinfo{volume}{B178}},
  \bibinfo{pages}{416} (\bibinfo{year}{1986}).

\bibitem[{\citenamefont{Korthals~Altes}(1994)}]{KorthalsAltes:1993ca}
\bibinfo{author}{\bibfnamefont{C.~P.} \bibnamefont{Korthals~Altes}},
  \bibinfo{journal}{Nucl. Phys.} \textbf{\bibinfo{volume}{B420}},
  \bibinfo{pages}{637} (\bibinfo{year}{1994}), \eprint{hep-th/9310195}.

\bibitem[{\citenamefont{Pisarski}(2002)}]{Pisarski:2002ji}
\bibinfo{author}{\bibfnamefont{R.~D.} \bibnamefont{Pisarski}}
  (\bibinfo{year}{2002}), \eprint{hep-ph/0203271}.

\bibitem[{\citenamefont{Gava and Jengo}(1981)}]{Gava:1981qd}
\bibinfo{author}{\bibfnamefont{E.}~\bibnamefont{Gava}} \bibnamefont{and}
  \bibinfo{author}{\bibfnamefont{R.}~\bibnamefont{Jengo}},
  \bibinfo{journal}{Phys. Lett.} \textbf{\bibinfo{volume}{B105}},
  \bibinfo{pages}{285} (\bibinfo{year}{1981}).

\bibitem[{\citenamefont{Kaczmarek et~al.}(2002)\citenamefont{Kaczmarek, Karsch,
  Petreczky, and Zantow}}]{Kaczmarek:2002mc}
\bibinfo{author}{\bibfnamefont{O.}~\bibnamefont{Kaczmarek}},
  \bibinfo{author}{\bibfnamefont{F.}~\bibnamefont{Karsch}},
  \bibinfo{author}{\bibfnamefont{P.}~\bibnamefont{Petreczky}},
  \bibnamefont{and} \bibinfo{author}{\bibfnamefont{F.}~\bibnamefont{Zantow}},
  \bibinfo{journal}{Phys. Lett.} \textbf{\bibinfo{volume}{B543}},
  \bibinfo{pages}{41} (\bibinfo{year}{2002}), \eprint{hep-lat/0207002}.

\bibitem[{\citenamefont{Necco and Sommer}(2002)}]{Necco:2001xg}
\bibinfo{author}{\bibfnamefont{S.}~\bibnamefont{Necco}} \bibnamefont{and}
  \bibinfo{author}{\bibfnamefont{R.}~\bibnamefont{Sommer}},
  \bibinfo{journal}{Nucl. Phys.} \textbf{\bibinfo{volume}{B622}},
  \bibinfo{pages}{328} (\bibinfo{year}{2002}), \eprint{hep-lat/0108008}.

\bibitem[{\citenamefont{Chishtie and Elias}(2001)}]{Chishtie:2001mf}
\bibinfo{author}{\bibfnamefont{F.~A.} \bibnamefont{Chishtie}} \bibnamefont{and}
  \bibinfo{author}{\bibfnamefont{V.}~\bibnamefont{Elias}},
  \bibinfo{journal}{Phys. Lett.} \textbf{\bibinfo{volume}{B521}},
  \bibinfo{pages}{434} (\bibinfo{year}{2001}), \eprint{hep-ph/0107052}.

\bibitem[{\citenamefont{Pineda}(2001)}]{Pineda:2001zq}
\bibinfo{author}{\bibfnamefont{A.}~\bibnamefont{Pineda}},
  \bibinfo{journal}{JHEP} \textbf{\bibinfo{volume}{06}}, \bibinfo{pages}{022}
  (\bibinfo{year}{2001}), \eprint{hep-ph/0105008}.

\bibitem[{\citenamefont{Cvetic}(2004)}]{Cvetic:2003wk}
\bibinfo{author}{\bibfnamefont{G.}~\bibnamefont{Cvetic}}, \bibinfo{journal}{J.
  Phys.} \textbf{\bibinfo{volume}{G30}}, \bibinfo{pages}{863}
  (\bibinfo{year}{2004}), \eprint{hep-ph/0309262}.

\bibitem[{\citenamefont{Hoang et~al.}(1999)\citenamefont{Hoang, Smith, Stelzer,
  and Willenbrock}}]{Hoang:1998nz}
\bibinfo{author}{\bibfnamefont{A.~H.} \bibnamefont{Hoang}},
  \bibinfo{author}{\bibfnamefont{M.~C.} \bibnamefont{Smith}},
  \bibinfo{author}{\bibfnamefont{T.}~\bibnamefont{Stelzer}}, \bibnamefont{and}
  \bibinfo{author}{\bibfnamefont{S.}~\bibnamefont{Willenbrock}},
  \bibinfo{journal}{Phys. Rev.} \textbf{\bibinfo{volume}{D59}},
  \bibinfo{pages}{114014} (\bibinfo{year}{1999}), \eprint{hep-ph/9804227}.

\bibitem[{\citenamefont{Beneke}(1998)}]{Beneke:1998rk}
\bibinfo{author}{\bibfnamefont{M.}~\bibnamefont{Beneke}},
  \bibinfo{journal}{Phys. Lett.} \textbf{\bibinfo{volume}{B434}},
  \bibinfo{pages}{115} (\bibinfo{year}{1998}), \eprint{hep-ph/9804241}.

\bibitem[{\citenamefont{Brambilla et~al.}(1999)\citenamefont{Brambilla, Pineda,
  Soto, and Vairo}}]{Brambilla:1999qa}
\bibinfo{author}{\bibfnamefont{N.}~\bibnamefont{Brambilla}},
  \bibinfo{author}{\bibfnamefont{A.}~\bibnamefont{Pineda}},
  \bibinfo{author}{\bibfnamefont{J.}~\bibnamefont{Soto}}, \bibnamefont{and}
  \bibinfo{author}{\bibfnamefont{A.}~\bibnamefont{Vairo}},
  \bibinfo{journal}{Phys. Rev.} \textbf{\bibinfo{volume}{D60}},
  \bibinfo{pages}{091502} (\bibinfo{year}{1999}), \eprint{hep-ph/9903355}.

\bibitem[{\citenamefont{Bali}(2001)}]{Bali:2000gf}
\bibinfo{author}{\bibfnamefont{G.~S.} \bibnamefont{Bali}},
  \bibinfo{journal}{Phys. Rept.} \textbf{\bibinfo{volume}{343}},
  \bibinfo{pages}{1} (\bibinfo{year}{2001}), \eprint{hep-ph/0001312}.

\bibitem[{\citenamefont{Sumino}(2003)}]{Sumino:2003yp}
\bibinfo{author}{\bibfnamefont{Y.}~\bibnamefont{Sumino}},
  \bibinfo{journal}{Phys. Lett.} \textbf{\bibinfo{volume}{B571}},
  \bibinfo{pages}{173} (\bibinfo{year}{2003}), \eprint{hep-ph/0303120}.

\bibitem[{\citenamefont{Andreev and Zakharov}(2006)}]{Andreev:2006nw}
\bibinfo{author}{\bibfnamefont{O.}~\bibnamefont{Andreev}} \bibnamefont{and}
  \bibinfo{author}{\bibfnamefont{V.~I.} \bibnamefont{Zakharov}}
  (\bibinfo{year}{2006}), \eprint{hep-ph/0611304}.

\bibitem[{\citenamefont{Wilson}(1969)}]{Wilson:1969zs}
\bibinfo{author}{\bibfnamefont{K.~G.} \bibnamefont{Wilson}},
  \bibinfo{journal}{Phys. Rev.} \textbf{\bibinfo{volume}{179}},
  \bibinfo{pages}{1499} (\bibinfo{year}{1969}).

\bibitem[{\citenamefont{Meg{\'\i}as
  et~al.}(2006{\natexlab{a}})\citenamefont{Meg{\'\i}as, Ruiz~Arriola, and
  Salcedo}}]{Megias:2005ve}
\bibinfo{author}{\bibfnamefont{E.}~\bibnamefont{Meg{\'\i}as}},
  \bibinfo{author}{\bibfnamefont{E.}~\bibnamefont{Ruiz~Arriola}},
  \bibnamefont{and} \bibinfo{author}{\bibfnamefont{L.~L.}
  \bibnamefont{Salcedo}}, \bibinfo{journal}{JHEP}
  \textbf{\bibinfo{volume}{01}}, \bibinfo{pages}{073}
  (\bibinfo{year}{2006}{\natexlab{a}}), \eprint{hep-ph/0505215}.

\bibitem[{\citenamefont{Kondo}(2001)}]{Kondo:2001nq}
\bibinfo{author}{\bibfnamefont{K.-I.} \bibnamefont{Kondo}},
  \bibinfo{journal}{Phys. Lett.} \textbf{\bibinfo{volume}{B514}},
  \bibinfo{pages}{335} (\bibinfo{year}{2001}), \eprint{hep-th/0105299}.

\bibitem[{\citenamefont{Meg{\'\i}as
  et~al.}(2006{\natexlab{b}})\citenamefont{Meg{\'\i}as, Arriola, and
  Salcedo}}]{Megias:2005pe}
\bibinfo{author}{\bibfnamefont{E.}~\bibnamefont{Meg{\'\i}as}},
  \bibinfo{author}{\bibfnamefont{E.~R.} \bibnamefont{Arriola}},
  \bibnamefont{and} \bibinfo{author}{\bibfnamefont{L.~L.}
  \bibnamefont{Salcedo}}, \bibinfo{journal}{Rom. Rep. Phys.}
  \textbf{\bibinfo{volume}{58}}, \bibinfo{pages}{081}
  (\bibinfo{year}{2006}{\natexlab{b}}), \eprint{hep-ph/0510114}.

\bibitem[{\citenamefont{Nadkarni}(1983)}]{Nadkarni:1983kb}
\bibinfo{author}{\bibfnamefont{S.}~\bibnamefont{Nadkarni}},
  \bibinfo{journal}{Phys. Rev.} \textbf{\bibinfo{volume}{D27}},
  \bibinfo{pages}{917} (\bibinfo{year}{1983}).

\bibitem[{\citenamefont{Ginsparg}(1980)}]{Ginsparg:1980ef}
\bibinfo{author}{\bibfnamefont{P.}~\bibnamefont{Ginsparg}},
  \bibinfo{journal}{Nucl. Phys.} \textbf{\bibinfo{volume}{B170}},
  \bibinfo{pages}{388} (\bibinfo{year}{1980}).

\bibitem[{\citenamefont{Appelquist and Pisarski}(1981)}]{Appelquist:1981vg}
\bibinfo{author}{\bibfnamefont{T.}~\bibnamefont{Appelquist}} \bibnamefont{and}
  \bibinfo{author}{\bibfnamefont{R.~D.} \bibnamefont{Pisarski}},
  \bibinfo{journal}{Phys. Rev.} \textbf{\bibinfo{volume}{D23}},
  \bibinfo{pages}{2305} (\bibinfo{year}{1981}).

\bibitem[{\citenamefont{Braaten and Nieto}(1996)}]{Braaten:1996jr}
\bibinfo{author}{\bibfnamefont{E.}~\bibnamefont{Braaten}} \bibnamefont{and}
  \bibinfo{author}{\bibfnamefont{A.}~\bibnamefont{Nieto}},
  \bibinfo{journal}{Phys. Rev.} \textbf{\bibinfo{volume}{D53}},
  \bibinfo{pages}{3421} (\bibinfo{year}{1996}), \eprint{hep-ph/9510408}.

\bibitem[{\citenamefont{Shaposhnikov}(1996)}]{Shaposhnikov:1996th}
\bibinfo{author}{\bibfnamefont{M.~E.} \bibnamefont{Shaposhnikov}}
  (\bibinfo{year}{1996}), \bibinfo{note}{effective theories and fundamental
  interactions, Erice 1996, 360-383}, \eprint{hep-ph/9610247}.

\bibitem[{\citenamefont{Meg{\'\i}as
  et~al.}(2006{\natexlab{c}})\citenamefont{Meg{\'\i}as, Arriola, and
  Salcedo}}]{Megias:2006ke}
\bibinfo{author}{\bibfnamefont{E.}~\bibnamefont{Meg{\'\i}as}},
  \bibinfo{author}{\bibfnamefont{E.~R.} \bibnamefont{Arriola}},
  \bibnamefont{and} \bibinfo{author}{\bibfnamefont{L.~L.}
  \bibnamefont{Salcedo}} (\bibinfo{year}{2006}{\natexlab{c}}),
  \eprint{hep-ph/0610163}.

\bibitem[{\citenamefont{Dumitru et~al.}(2004)\citenamefont{Dumitru, Hatta,
  Lenaghan, Orginos, and Pisarski}}]{Dumitru:2003hp}
\bibinfo{author}{\bibfnamefont{A.}~\bibnamefont{Dumitru}},
  \bibinfo{author}{\bibfnamefont{Y.}~\bibnamefont{Hatta}},
  \bibinfo{author}{\bibfnamefont{J.}~\bibnamefont{Lenaghan}},
  \bibinfo{author}{\bibfnamefont{K.}~\bibnamefont{Orginos}}, \bibnamefont{and}
  \bibinfo{author}{\bibfnamefont{R.~D.} \bibnamefont{Pisarski}},
  \bibinfo{journal}{Phys. Rev.} \textbf{\bibinfo{volume}{D70}},
  \bibinfo{pages}{034511} (\bibinfo{year}{2004}), \eprint{hep-th/0311223}.

\bibitem[{\citenamefont{Kaczmarek and
  Zantow}(2005{\natexlab{a}})}]{Kaczmarek:2005ui}
\bibinfo{author}{\bibfnamefont{O.}~\bibnamefont{Kaczmarek}} \bibnamefont{and}
  \bibinfo{author}{\bibfnamefont{F.}~\bibnamefont{Zantow}},
  \bibinfo{journal}{Phys. Rev.} \textbf{\bibinfo{volume}{D71}},
  \bibinfo{pages}{114510} (\bibinfo{year}{2005}{\natexlab{a}}),
  \eprint{hep-lat/0503017}.

\bibitem[{\citenamefont{Pascual and Tarrach}(1984)}]{Pascual:1984zb}
\bibinfo{author}{\bibfnamefont{P.}~\bibnamefont{Pascual}} \bibnamefont{and}
  \bibinfo{author}{\bibfnamefont{R.}~\bibnamefont{Tarrach}},
  \bibinfo{journal}{Lect. Notes Phys.} \textbf{\bibinfo{volume}{194}},
  \bibinfo{pages}{1} (\bibinfo{year}{1984}).

\bibitem[{\citenamefont{Shifman et~al.}(1979)\citenamefont{Shifman, Vainshtein,
  and Zakharov}}]{Shifman:1978bx}
\bibinfo{author}{\bibfnamefont{M.~A.} \bibnamefont{Shifman}},
  \bibinfo{author}{\bibfnamefont{A.~I.} \bibnamefont{Vainshtein}},
  \bibnamefont{and} \bibinfo{author}{\bibfnamefont{V.~I.}
  \bibnamefont{Zakharov}}, \bibinfo{journal}{Nucl. Phys.}
  \textbf{\bibinfo{volume}{B147}}, \bibinfo{pages}{385} (\bibinfo{year}{1979}).

\bibitem[{\citenamefont{Narison}(1989)}]{Narison:1989aq}
\bibinfo{author}{\bibfnamefont{S.}~\bibnamefont{Narison}},
  \bibinfo{journal}{World Sci. Lect. Notes Phys.}
  \textbf{\bibinfo{volume}{26}}, \bibinfo{pages}{1} (\bibinfo{year}{1989}).

\bibitem[{\citenamefont{Lavelle and Schaden}(1988)}]{Lavelle:1988eg}
\bibinfo{author}{\bibfnamefont{M.~J.} \bibnamefont{Lavelle}} \bibnamefont{and}
  \bibinfo{author}{\bibfnamefont{M.}~\bibnamefont{Schaden}},
  \bibinfo{journal}{Phys. Lett.} \textbf{\bibinfo{volume}{B208}},
  \bibinfo{pages}{297} (\bibinfo{year}{1988}).

\bibitem[{\citenamefont{Chetyrkin et~al.}(1999)\citenamefont{Chetyrkin,
  Narison, and Zakharov}}]{Chetyrkin:1998yr}
\bibinfo{author}{\bibfnamefont{K.~G.} \bibnamefont{Chetyrkin}},
  \bibinfo{author}{\bibfnamefont{S.}~\bibnamefont{Narison}}, \bibnamefont{and}
  \bibinfo{author}{\bibfnamefont{V.~I.} \bibnamefont{Zakharov}},
  \bibinfo{journal}{Nucl. Phys.} \textbf{\bibinfo{volume}{B550}},
  \bibinfo{pages}{353} (\bibinfo{year}{1999}), \eprint{hep-ph/9811275}.

\bibitem[{\citenamefont{Boucaud et~al.}(2001)}]{Boucaud:2001st}
\bibinfo{author}{\bibfnamefont{P.}~\bibnamefont{Boucaud}} \bibnamefont{et~al.},
  \bibinfo{journal}{Phys. Rev.} \textbf{\bibinfo{volume}{D63}},
  \bibinfo{pages}{114003} (\bibinfo{year}{2001}), \eprint{hep-ph/0101302}.

\bibitem[{\citenamefont{Ruiz~Arriola et~al.}(2004)\citenamefont{Ruiz~Arriola,
  Bowman, and Broniowski}}]{RuizArriola:2004en}
\bibinfo{author}{\bibfnamefont{E.}~\bibnamefont{Ruiz~Arriola}},
  \bibinfo{author}{\bibfnamefont{P.~O.} \bibnamefont{Bowman}},
  \bibnamefont{and}
  \bibinfo{author}{\bibfnamefont{W.}~\bibnamefont{Broniowski}},
  \bibinfo{journal}{Phys. Rev.} \textbf{\bibinfo{volume}{D70}},
  \bibinfo{pages}{097505} (\bibinfo{year}{2004}), \eprint{hep-ph/0408309}.

\bibitem[{\citenamefont{Boucaud et~al.}(2005)}]{Boucaud:2005rm}
\bibinfo{author}{\bibfnamefont{P.}~\bibnamefont{Boucaud}} \bibnamefont{et~al.}
  (\bibinfo{year}{2005}), \eprint{hep-lat/0504017}.

\bibitem[{\citenamefont{Philipsen}(2002)}]{Philipsen:2002az}
\bibinfo{author}{\bibfnamefont{O.}~\bibnamefont{Philipsen}},
  \bibinfo{journal}{Phys. Lett.} \textbf{\bibinfo{volume}{B535}},
  \bibinfo{pages}{138} (\bibinfo{year}{2002}), \eprint{hep-lat/0203018}.

\bibitem[{\citenamefont{Kaczmarek et~al.}(2004)\citenamefont{Kaczmarek, Karsch,
  Zantow, and Petreczky}}]{Kaczmarek:2004gv}
\bibinfo{author}{\bibfnamefont{O.}~\bibnamefont{Kaczmarek}},
  \bibinfo{author}{\bibfnamefont{F.}~\bibnamefont{Karsch}},
  \bibinfo{author}{\bibfnamefont{F.}~\bibnamefont{Zantow}}, \bibnamefont{and}
  \bibinfo{author}{\bibfnamefont{P.}~\bibnamefont{Petreczky}},
  \bibinfo{journal}{Phys. Rev.} \textbf{\bibinfo{volume}{D70}},
  \bibinfo{pages}{074505} (\bibinfo{year}{2004}), \eprint{hep-lat/0406036}.

\bibitem[{\citenamefont{Anisovich et~al.}(2000)\citenamefont{Anisovich,
  Anisovich, and Sarantsev}}]{Anisovich:2000kx}
\bibinfo{author}{\bibfnamefont{A.~V.} \bibnamefont{Anisovich}},
  \bibinfo{author}{\bibfnamefont{V.~V.} \bibnamefont{Anisovich}},
  \bibnamefont{and} \bibinfo{author}{\bibfnamefont{A.~V.}
  \bibnamefont{Sarantsev}}, \bibinfo{journal}{Phys. Rev.}
  \textbf{\bibinfo{volume}{D62}}, \bibinfo{pages}{051502}
  (\bibinfo{year}{2000}), \eprint{hep-ph/0003113}.

\bibitem[{\citenamefont{Zantow}(2003)}]{Zantow:2003uh}
\bibinfo{author}{\bibfnamefont{F.}~\bibnamefont{Zantow}}
  (\bibinfo{year}{2003}), \eprint{hep-lat/0301014}.

\bibitem[{\citenamefont{Kaczmarek and
  Zantow}(2005{\natexlab{b}})}]{Kaczmarek:2005jy}
\bibinfo{author}{\bibfnamefont{O.}~\bibnamefont{Kaczmarek}} \bibnamefont{and}
  \bibinfo{author}{\bibfnamefont{F.}~\bibnamefont{Zantow}}
  (\bibinfo{year}{2005}{\natexlab{b}}), \eprint{hep-lat/0512031}.

\bibitem[{\citenamefont{Agasian and Simonov}(2006)}]{Agasian:2006ra}
\bibinfo{author}{\bibfnamefont{N.~O.} \bibnamefont{Agasian}} \bibnamefont{and}
  \bibinfo{author}{\bibfnamefont{Y.~A.} \bibnamefont{Simonov}},
  \bibinfo{journal}{Phys. Lett.} \textbf{\bibinfo{volume}{B639}},
  \bibinfo{pages}{82} (\bibinfo{year}{2006}), \eprint{hep-ph/0604004}.

\end{thebibliography}

\end{document}